\UseRawInputEncoding
\pdfoutput=1
\documentclass{article}
\usepackage{spconf,amsmath,epsfig}
\usepackage{color}
\usepackage{enumitem}
\usepackage{amssymb}
\usepackage{booktabs}
\usepackage{multirow}
\usepackage{pifont}
\usepackage{bbding}
\usepackage{threeparttable}
\usepackage{subfigure}

\definecolor{holder}{rgb}{0.87,0.78,1.0}
\definecolor{duplicate}{rgb}{0.52,0.85,0.81}
\definecolor{meaning}{rgb}{0.75,0.75,0.75}
\definecolor{smooth}{rgb}{0.52,0.52,0.52} 
\definecolor{catching}{rgb}{1.0,0.5,0.0}

\newcommand{\ours}{{SRCR}}

\let\OLDthebibliography\thebibliography
\renewcommand\thebibliography[1]{
  \OLDthebibliography{#1}
  \setlength{\parskip}{0pt}
  \setlength{\itemsep}{0pt plus 0.3ex}
}

\begin{document}\sloppy

\def\x{{\mathbf x}}
\def\L{{\cal L}}

\title{Structure-Aware Residual-Center Representation for \\ Self-Supervised Open-Set 3D Cross-Modal Retrieval}
\name{Yang Xu$^1$, Yifan Feng$^1$, Yu Jiang$^{2\dagger}$\thanks{$^{\dagger}$Corresponding author: Yu Jiang}}
\address{$^1$ School of Software, Tsinghua University;\\ $^2$ College of Computer Science and Technology, Jilin University;\\
\{xuyang9610, evanfeng97\}@gmail.com; jiangyu2011@jlu.edu.cn}
\maketitle

\begin{abstract}
Existing methods of 3D cross-modal retrieval heavily lean on category distribution priors within the training set, which diminishes their efficacy when tasked with unseen categories under open-set environments. To tackle this problem, we propose the Structure-Aware Residual-Center Representation (\ours) framework for self-supervised open-set 3D cross-modal retrieval. To address the center deviation due to category distribution differences, we utilize the Residual-Center Embedding (RCE) for each object by nested auto-encoders, rather than directly mapping them to the modality or category centers. Besides, we perform the Hierarchical Structure Learning (HSL) approach to leverage the high-order correlations among objects for generalization, by constructing a heterogeneous hypergraph structure based on hierarchical inter-modality, intra-object, and implicit-category correlations. Extensive experiments and ablation studies on four benchmarks demonstrate the superiority of our proposed framework compared to state-of-the-art methods.
\end{abstract}


\section{Introduction}
\label{section_1}
The proliferation of multimedia data on the Internet, including videos, images, text, and more, has sparked growing interest within the community in the field of cross-modal retrieval tasks. Among them, 3D cross-modal retrieval (3DCMR) has garnered growing attention due to the inherent diverse modalities of 3D data~\cite{jing2021cross} and its relevance across crucial domains such as robotics, medicine, and other significant fields.

Typical 3D cross-modal retrieval task aims to retrieve 3D data from one modality given queries from different modalities. To address the heterogeneity gap~\cite{wang2017adversarial} from different modalities, a widely adopted strategy of 3D cross-modal retrieval methods is to seek a function that maps data samples from diverse modalities into a unified global representation space~\cite{jing2021cross,feng2023rono}, which is called the \textit{center}. 

\begin{figure}[!t]
\centering
\includegraphics[bb=0 0 538.56 311.76, width=\linewidth]{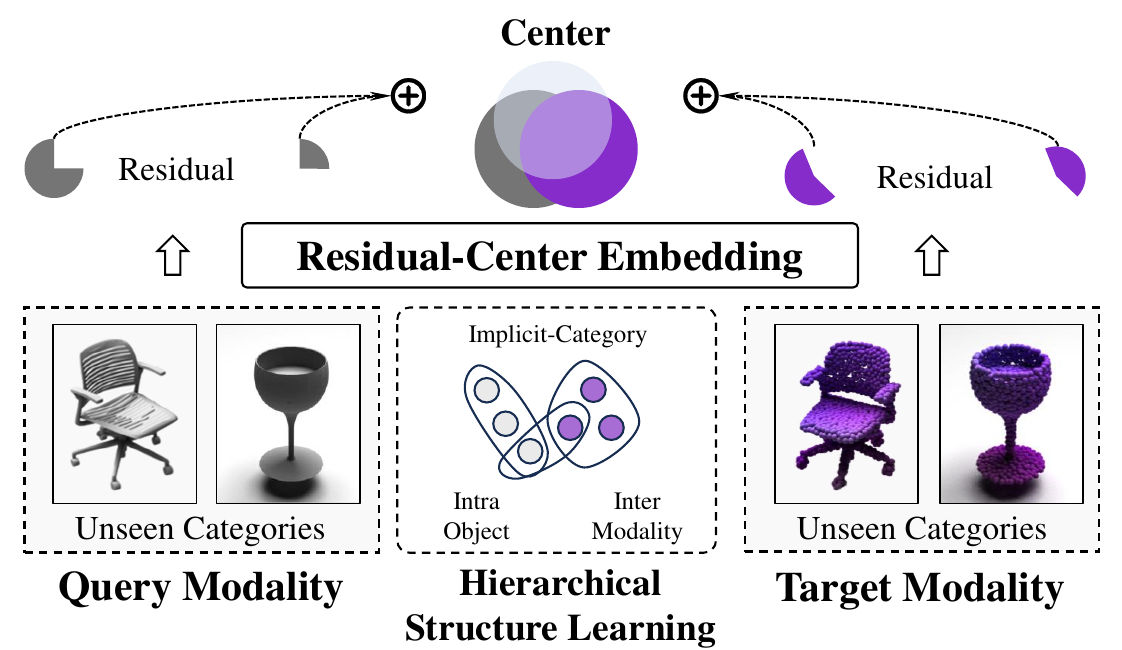}       
\caption{Illustration of the proposed {\ours}. Given 3D objects of unseen categories represented by different modalities, our method generates the residual-center embeddings for each modality of each object. Then unified center representations are generated via hierarchical structure learning for cross-modal retrieval with unseen categories generalization.}
\label{figure_1_1}
\end{figure}

Current methods of constructing such mapping can be broadly categorized into two approaches. One straightforward solution for this task is to construct complex nonlinear transformations~\cite{andrew2013deep,wang2015deep} that map two types of pre-trained features into a shared space. The alternative approach employs adversarial loss to learn category-related central embedding~\cite{jing2021cross,feng2023rono} through end-to-end training. However, both methods exhibit a pronounced dependence on the prior distribution of category spaces within the training sets, which leads to substantial representational biases when confronted with objects of unseen categories. Furthermore, dependency on training labels in adversarial loss also complicates the deployment of 3D cross-modal retrieval.

To overcome the aforementioned challenges, we propose the Structure-Aware Residual-Center Representation (\ours) framework for the self-supervised open-set 3D cross-modal retrieval task, as shown in Fig.~\ref{figure_1_1}. On one hand, to overcome the center deviation due to category distribution differences, we utilize the residual-center embedding for each object by nested auto-encoders, rather than directly mapping them to the modality or category center. On the other hand, we perform a hierarchical structure learning approach to utilize the high-order correlations among objects for generalization, by constructing a heterogeneous hypergraph structure based on hierarchical intra-modality, inter-object, and implicit-category correlations. Our contributions are summarized as follows:
\begin{itemize}[itemsep=0pt]
\item{We introduce a practical open-set setting for 3D cross-modal retrieval and generate four datasets for benchmarking of downstream 3D cross-modal tasks.}
\item{We propose the Structure-Aware Residual-Center Representation (\ours) framework for the open-set 3D cross-modal retrieval, including the Residual-Center Embedding (RCE) and Hierarchical Structure Learning (HSL) modules, which are designed to overcome the modality diversion accentuated by unseen categories distribution.}
\item{We propose a hierarchical hypergraph structure to capture the high-order correlations among objects, under the guidance of hierarchical inter-modality, intra-object, and implicit-category correlations.}
\item{The proposed framework significantly outperforms the state-of-the-art 3D cross-modal retrieval methods under the open-set setting.}
\end{itemize}
\section{Related Work}
\label{section_2}
\subsection{Cross-Modal Retrieval}
Existing methods usually construct a mapping function into a unified common space to overcome the heterogeneity gap~\cite{wang2017adversarial} from different modalities. These approaches could be roughly classified into projection-based~\cite{andrew2013deep} and discrimination-based~\cite{jing2021cross,feng2023rono} methods. While such methods excel under the closed-set assumption, their reliance on training category distribution limits their generalization in real-world, open-set environments.

\subsection{Open-Environment Learning}
Most current methods of open-environment learning are designed for open-set recognition~\cite{vaze2021open,chen2021adversarial}, which is usually used to detect whether the sample belongs to the seen categories or not. While some methods have succeeded in open-set 3D multi-modal retrieval~\cite{feng2023hypergraph}, the complexity and inherent disparities between different modalities still present considerable challenges in open-set cross-modal retrieval.

\section{Methodology}
\label{section_3}
\subsection{Problem Setup}
\label{section_3_1}
Given $N$ 3D objects $\{o_i\} = \{ o^r_i \}_{r=1}^{M}$ represented by $M$ modalities, the goal of 3D cross-modal retrieval (3DCMR) is to develop a model using the training set $\mathcal{D}_{trn} = \{(o_i, y_i)\}^{L}_{i=1}$, and then employ it to identify similar objects from the query set $\mathcal{D}_{q} = \{(o^q_i, \hat{y}_i)\}^{Q}_{i=1}$ to the target set $\mathcal{D}_{t} = \{(o^t_i, \hat{y}_i)\}^{T}_{i=1}$, where the query and target objects are represented in different modalities ($t \neq q$). Here, $L$, $Q$, and $T$ denote the number of samples in the training, query, and target set, respectively. The query set and target set are from testing set $\mathcal{D}_{tes} = \{\mathcal{D}_{q},\mathcal{D}_{t} \}$, $y_i \in \mathcal{Y} = \{c_j\}^Y_{j=1}$ and $\hat{y}_i \in \mathcal{\hat{Y}} = \{\hat{c}_j\}^{\hat{Y}}_{j=1}$ denote the category space of the training set and target set, where $Y$ and $\hat{Y}$ are the numbers of categories in the training and testing sets, respectively.

Traditional 3DCMR task is based on the close-set assumption, which means that in the testing set $\mathcal{D}_{tes} = \{ \mathcal{D}_{q},\mathcal{D}_{t}\}$, all categories of objects in the testing set have been seen in the training set $\mathcal{D}_{trn}$. The category spaces of the training set and testing set are the same indicating $\mathcal{Y} = \mathcal{\hat{Y}}$.

Different from the traditional closed-set assumption, we consider a more practical condition that the testing set consists entirely of categories not encountered in the training set. We term this task as \textit{Open-Set 3D Cross-Modal Retrieval}. Under this circumstance, $\mathcal{D}_{trn}$ and $\mathcal{D}_{tes}$ have their individual distributions, which means $\mathcal{Y} \neq \mathcal{\hat{Y}}$. This task seeks to minimize the expected risk:
\begin{equation}
\begin{split}
f^* = \mathop{argmin}\limits_{f\in \mathcal{H}} &\mathbb{E}_{( D_i, D_j) \sim (\mathcal{D}_q, \mathcal{D}_t)} \left[\mathbb{I}(\hat{y}_i \neq \hat{y}_j) e^{-\mathbb{D}(f(o^q_i), f(o^t_j))} \right.\\ &\left. +\mathbb{I}(\hat{y}_i = \hat{y}_j) (1 - e^{-\mathbb{D}(f(o^q_i), f(o^t_j))} ) \right]
\end{split},
\label{equation_3_2}
\end{equation}

\noindent where $D_i = (o^q_i, \hat{y}_i)$ and $D_j = (o^t_j, \hat{y}_j)$ are samples drawn from the query set $\mathcal{D}_q$ and target set $\mathcal{D}_t$. $\mathbb{I}(\cdot)$ is the indicator function, which returns $1$ if the expression is true and $0$ otherwise. $f := o^r_i \rightarrow z_i$ is the function that maps the 3D object $o^r_i$ represented in different modalities into the same embedding $z_i \in \mathbb{R}^d$. $\mathcal{H}$ is the hypothesis space of function $f(\cdot)$ and $\mathbb{D}(z_i, z_j)$ is a distance metric function.

\begin{figure*}[!t]
\centering
\includegraphics[bb=0 0 850.32 283.56, width=\linewidth]{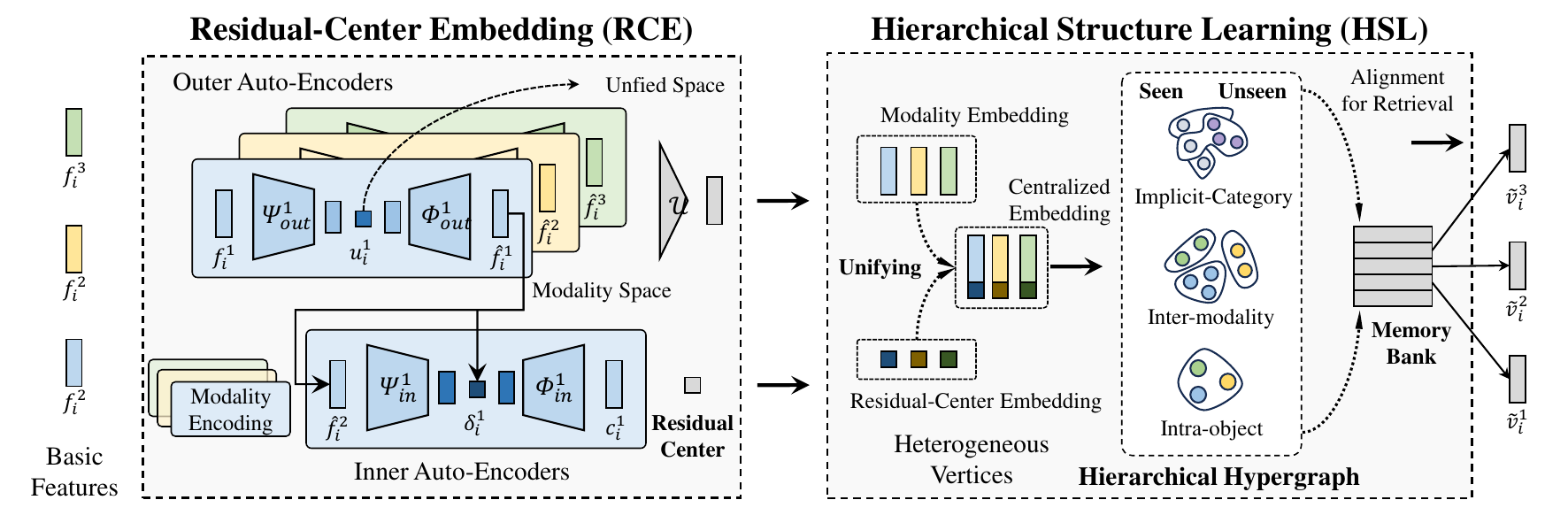}    
\caption{An overview of the proposed structure-aware residual-center representation framework ({\ours}). Our framework comprises two main modules: Residual-Center Embedding (RCE) and Hierarchical Structure Learning (HSL), which are used for residual embedding generation and structure-aware feature alignment, respectively.}
\label{figure_3_1}
\end{figure*}

\subsection{Framework Architecture}
\label{section_3_2}
The architecture of {\ours}, as illustrated in Fig.~\ref{figure_3_1}, is composed of two modules: \textit{Residual-Center Embedding (RCE)} and \textit{Hierarchical Structure Learning (HSL)}. Given basic features of different modalities extracted by common-used networks. The Residual-Center Embedding module is designed to generate the residual center embeddings for each object, rather than directly mapping them to the modality or category center. Then, in the Hierarchical Structure Learning stage, the hierarchical hypergraph structure is constructed based on the inter-modality, intra-object, and implicit-category correlations. Guided by this structure, the combination of hypergraph convolution and memory bank effectively leverages the high-order correlations between seen and unseen categories and different modalities. Finally, the aligned embedding of each modality is generated for the cross-modal retrieval or other downstream tasks.

\subsection{Residual-Center Embedding}
\label{section_3_3}
In order to improve category generalization while projecting into the unified space, the residual-center embedding module is developed. Specifically, the RCE consists of two nested auto-encoders and takes the basic features of different modalities as input. The outer auto-encoder $\mathcal{A}^{r}_{out}$ encodes the basic features into a latent space and pulls them together into a unified embedding. The inner auto-encoder $\mathcal{A}^{r}_{in}$ encodes the modality embeddings from the outer auto-encoder to the residual space, which transforms the embedding between the modality space and the unified space.

\subsubsection{Residual Learning}
Given $N$ 3D objects $\{o_i\}^N_{i=1}$ and basic features $\{f^r_i\}^M_{r=1}$ of each object. As shown in Fig.~\ref{figure_3_1}, the outer auto-encoder $\mathcal{A}^{r}_{out} = \{ \Psi^{r}_{out}, \Phi^{r}_{out} \}_{r=1}^{M}$ compresses the basic features into a unified space $\mathbb{S}_u$ and does the reverse reconstruction, for better representation, which can be defined as follows:
\begin{equation}
\left\{
\begin{aligned}
u^r_i = & \Psi^{r}_{out} (f^r_i) \\
\hat{f}^r_i = & \Phi^{r}_{out} (u^r_i) \\
\end{aligned}
\right. ,
\label{equation_3_3_1}
\end{equation}

\noindent where $\Psi^{r}_{out} := \mathbb{S}_r \rightarrow \mathbb{S}_u$ is the encoder that maps the $r$-th modality space $\mathbb{S}_r$ into the unified space $\mathbb{S}_u$, and $\Phi^{r}_{out} := \mathbb{S}_u \rightarrow \mathbb{S}_r$ is the decoder that maps the features from unified space $\mathbb{S}_u$ back to $\mathbb{S}_r$. $u^r_i \in \mathbb{R}^{d_u}$ and $\hat{f}^n_i \in \mathbb{R}^{d_0}$ denotes the compressed features and reconstruction features of each modality. 

An aggregation function $\mathcal{U}$ are adopted to generate the unified embedding $u_i = \mathcal{U}(\{u^r_i \}_{r=1}^{M}), u_i \in \mathbb{R}^{d_u}$ of object $o_i$, which are treated as the semantic center. The inner auto-encoder $\mathcal{A}^{r}_{in} = \{ \Psi^{r}_{in}, \Phi^{r}_{in} \}_{r=1}^{M}$ aims to generate the semantic center $u_i$ of each object and the residual-center embedding between each modality $\hat{f}^n_i$. Specifically, we construct learnable parameter encodings $e^r \in \mathbb{R}^{d_u}$ for each modality, and $\mathcal{A}^{r}_{in}$ takes them aligned with $\hat{f}^r_i$ to get middle embedding:
\begin{equation}
\left\{
\begin{aligned}
& \delta^r_i = \Psi^{r}_{in} ( \hat{f}^r_i + e^r ) \\
& c^r_i = \Phi^{r}_{in} (\hat{f}^r_i + \delta^r_i)
\end{aligned}
\right. ,
\label{equation_3_3_2}
\end{equation}

\noindent where $\Psi^{r}_{in}$ and $\Phi^{r}_{in}$ denote the encoder and decoder map function between modality and residual space, $\delta^r_i$ denotes the residual-center embedding of $r$-th modality of object $o_i$.

\subsubsection{Loss Function for RCE}
To get a better representation of modality embedding and residual-center embedding, the Residual-Center Loss $\mathcal{L}_{rc}$ and Cross-Reconstruction Loss $\mathcal{L}_{cr}$ are adopted here. The constraints for each loss are derived from different modalities data of the same object, rather than class labels.

\noindent \textbf{Residual-Center Loss.} The loss is designed to pull the distance among the estimated embeddings $\{u^r_i \}_{r=1}^{M}$ from different modalities closer, which is defined as follows:
\begin{equation}
\mathcal{L}_{rc} = \frac{1}{M} \sum\nolimits_{r=1}^M (\lVert u^r_i - u_i \rVert_2 + \lVert c^r_i - u_i \rVert_2),
\label{equation_3_3_3}
\end{equation}

\noindent where $\lVert \cdot \rVert_2$ is the $\mathcal{L}_2$ norm function.

\noindent \textbf{Cross-Reconstrution Loss.} To promote the generalization ability of the RCE, we propose the cross-reconstruction Loss. Motivated by~\cite{feng2014cross}, the $\mathcal{L}_{cr}$ are defined as the distance of results by exchanging decoder in the inner auto-encoder.
\begin{equation}
\mathcal{L}_{cr} = \frac{1}{M(M-1)} \sum_{k=1}^M \sum_{l\neq k} (\lVert \Phi^{l}_{in} ( \Psi^{k}_{in} ( \hat{f}^k_i + \delta^k_i )) - c^k_i \rVert_2) ,
\label{equation_3_3_4}
\end{equation}

\noindent where $\lVert \cdot \rVert_2$ is the $\mathcal{L}_2$ norm function.

\noindent \textbf{Joint Optimization.} In the residual-center embedding stage, the overall loss function is given combined Eq.~\ref{equation_3_3_3} and Eq.~\ref{equation_3_3_4}:
\begin{equation}
\mathcal{L}_{RCE} = \alpha \mathcal{L}_{rc} + (1-\alpha) \mathcal{L}_{cr},
\label{equation_3_3_5}
\end{equation}
where $\alpha$ is the hyper-parameter for trade-off.

\subsection{Hierarchical Structure Learning}
\label{section_3_4}
Although the RCE module generates the residual-center embedding of different modalities, the distribution gaps between seen and unseen categories still affect the retrieval under the open-set setting. As shown in Fig.~\ref{figure_3_1}, we proposed the hierarchical structure learning module for generalization across modalities and categories. Specifically, the hierarchical hypergraph is constructed to capture the hierarchical correlations. Then, the hypergraph convolution and memory bank are adopted for embedding smoothing and distilling.

\subsubsection{Hierarchical Hypergraph Construction}
We adopt a hierarchical hypergraph to take the most advantage of high-order correlations between modalities, objects, and categories. A hypergraph can be represented as $\mathcal{G} = \{ \mathcal{V}, \mathcal{E}\}$, where $\mathcal{V}$ and $\mathcal{E}$ are the vertex set and the hyperedge set, respectively.  

\noindent \textbf{Heterogeneous Vertices.} For the vertices, we first construct the centralized embedding of each modality by aligning $\hat{f}^r_i$ with the residual feature $\delta^r_i$, then we treat centralized embeddings of each object as the heterogeneous vertices.
\begin{equation}
\label{equation_3_4_1}
\left.
\begin{aligned}
&v^r_i = \tau \hat{f}^r_i + (1-\tau) \delta^r_i \\
&\mathcal{V} = \bigcup \nolimits_{r=1}^M \{v^r_i\}_{i=1}^N
\end{aligned}
\right. ,
\end{equation}

\noindent where $\tau$ denotes the hyper-parameters for centralized fusion, $\hat{f}^r_i$ and $\delta^r_i$ denote the modality embedding and residual-center embedding of object $o_i$ in $r$-th modality, $M$ and $N$ denote the number of modalities and object samples.

\noindent \textbf{Hierarchical Hyperedges.} The hierarchical hypergraph is composed of three types of hyperedges, including inter-modality, intra-object, and implicit-category, which can be defined as follows:
\begin{equation}
\label{equation_3_4_2}
\left.
\begin{aligned}
&\mathcal{E}_{m}=\{ \mathcal{M}_{v}(r) \mid r \in \{1,\cdots,M\} \}\\
&\mathcal{E}_{o}=\{ \mathcal{N}_{v}(i) \mid i \in \{1,\cdots,N\} \}\\
&\mathcal{E}_{c} = \{ \mathcal{N}_{\mathrm{KNN}_{k}}(v) \mid v \in \mathcal{V} \}
\end{aligned}
\right. ,
\end{equation}

\noindent where $\mathcal{M}_{v}(r)$ denotes the vertex subset that belong to the same modality $r$, $\mathcal{N}_{v}(i)$ denotes the vertex subset that belong to the same object $o_i$, and $\mathcal{N}_{\mathrm{KNN}_{k}}(v)$ denotes the k-nearest neighbors of vertex $v$.

In this way, $M$ inter-modality hyperedges, $N$ intra-object hyperedges and $M\times N$ implicit-category hyperedges are constructed. Finally, we combine these three hyperedge groups to get the total hyperedges: $\mathcal{E}= \mathcal{E}_{m} \cup \mathcal{E}_{o} \cup \mathcal{E}_{c}$.

\subsubsection{Hypergraph Convolution and Alignment}
To leverage the high-order correlation between objects and modalities, we utilize the hypergraph convolution~\cite{gao2022hgnn+} to smooth the embedding under the hierarchical structure, which is formulated as:
\begin{equation}
\tilde{\mathbf{V}} = \sigma\left( \mathbf{D}^{-\frac{1}{2}}_v \mathbf{H} \mathbf{W} \mathbf{D}^{-1}_e \mathbf{H}^\top \mathbf{D}^{-\frac{1}{2}}_v \mathbf{V} \mathbf{\Theta} \right) ,
\label{equation_3_4_3}
\end{equation}

\noindent where $\mathbf{H}$ denotes the incidence matrix of the hypergraph. $\mathbf{D}_v$ and $\mathbf{D}_e$ are the diagonal degree matrices for vertex and hyperedge, respectively.

After obtaining the structure-aware embedding $\tilde{v}^r_i$ of the 3D object $o^r_i$, we construct a memory bank $\mathcal{B}$ that contains $L$ invariant memory anchors. Following~\cite{feng2023hypergraph}, we compute the activation score for each memory anchor in the memory bank by $s^r_{ij} = \mathcal{D}_m(\tilde{v}^r_i, a^r_j)$, where $a^r_j$ denotes the anchor and $D_m(\cdot,\cdot)$ denotes the distance metric function. We rebuild the aligned embedding of each object by $z_i=\sum\nolimits_{j=1}^L \hat{s}^r_{ij}a^r_j, z^r_i \in \mathbb{R}^{d_z}$, where $\hat{s}^r_{ij}$ denotes the normlization of activation score.

\subsubsection{Loss Function for HSL}
To train the hypergraph convolution and learnable memory anchors under hierarchical structure, we adopt the \textbf{self-supervised} Memory Reconstruction Loss $\mathcal{L}_{mr}$ for HSL:
\begin{equation}
\mathcal{L}_{HSL} = \mathcal{L}_{mr} = \big\lVert \tilde{v}^r_i - z^r_i \big\rVert_2 ,
\label{equation_3_4_4}
\end{equation}

\noindent where $\lVert \cdot \rVert_2$ is the $\mathcal{L}_2$ norm function.
\begin{table*}[!htbp]
\begin{center}
\small
\caption{Experimental results of Image2Point retrieval on the OCAB, OCNT, OCES, and OCMN datasets.}
\scalebox{0.85}{
\begin{tabular}{c|ccc|ccc|ccc|ccc}
\toprule
\multirow{2}{*}{\textbf{Image2Point}} & \multicolumn{3}{c|}{\textbf{OCAB}} & \multicolumn{3}{c|}{\textbf{OCNT}} & \multicolumn{3}{c|}{\textbf{OCES}} & \multicolumn{3}{c}{\textbf{OCMN}} \\
 & \textbf{mAP}$\uparrow$ & \textbf{NDCG}$\uparrow$ & \textbf{ANMRR}$\downarrow$ & \textbf{mAP}$\uparrow$ & \textbf{NDCG}$\uparrow$ & \textbf{ANMRR}$\downarrow$ & \textbf{mAP}$\uparrow$ & \textbf{NDCG}$\uparrow$ & \textbf{ANMRR}$\downarrow$ & \textbf{mAP}$\uparrow$ & \textbf{NDCG}$\uparrow$ & \textbf{ANMRR}$\downarrow$ \\ \midrule
\textbf{SDML} &0.1489  & 0.1061  & 0.8824   & 0.0465 & 0.0316 & 0.9657  & 0.0942  & 0.0442  & 0.9486   & 0.0578 & 0.0248 & 0.9735 \\
\textbf{CMCL} &0.1702  & 0.1520  & 0.8565   & 0.0623  & 0.0332  & 0.9665   & 0.0991  & 0.0477  & 0.9444   & 0.1175  & 0.0917  & 0.9001 \\
\textbf{MMSAE} & 0.1218  & 0.0802  & 0.9093   & 0.0410  & 0.0191  & 0.9817   & 0.0810  & 0.0362  & 0.9567   & 0.0571  & 0.0235  & 0.9746 \\
\textbf{PROSER} &0.1119  & 0.0446  & 0.9386   & 0.0426  & 0.0171  & 0.9752   & 0.0968  & 0.0402  & 0.9641   & 0.0523  & 0.0133  & 0.9806  \\
\textbf{HGM$^2$R} & 0.1367  & 0.0925  & 0.8978   & 0.1812  & 0.1072  & 0.8184   & 0.2184  & 0.1126  & 0.8215   & 0.0988  & 0.0789  & 0.9282  \\
\midrule
\textbf{Ours} & \textbf{0.2220 } & \textbf{0.1714 } & \textbf{0.7947  } & \textbf{0.2861 } & \textbf{0.1585 } & \textbf{0.7292  } & \textbf{0.4004 } & \textbf{0.1835 } & \textbf{0.6378  } & \textbf{0.1549 } & \textbf{0.1488 } & \textbf{0.8625  } \\ 
\bottomrule
\end{tabular}
}
\label{table_4_2_1}
\end{center}
\end{table*}

\begin{table*}[!htbp]
\begin{center}
\small
\caption{Experimental results of Point2Image retrieval on the OCAB, OCNT, OCES, and OCMN datasets.}
\scalebox{0.85}{
\begin{tabular}{c|ccc|ccc|ccc|ccc}
\toprule
\multirow{2}{*}{\textbf{Point2Image}} & \multicolumn{3}{c|}{\textbf{OCAB}} & \multicolumn{3}{c|}{\textbf{OCNT}} & \multicolumn{3}{c|}{\textbf{OCES}} & \multicolumn{3}{c}{\textbf{OCMN}} \\
 & \textbf{mAP}$\uparrow$ & \textbf{NDCG}$\uparrow$ & \textbf{ANMRR}$\downarrow$ & \textbf{mAP}$\uparrow$ & \textbf{NDCG}$\uparrow$ & \textbf{ANMRR}$\downarrow$ & \textbf{mAP}$\uparrow$ & \textbf{NDCG}$\uparrow$ & \textbf{ANMRR}$\downarrow$ & \textbf{mAP}$\uparrow$ & \textbf{NDCG}$\uparrow$ & \textbf{ANMRR}$\downarrow$ \\ \midrule
\textbf{SDML} &0.1636 & 0.1367 & 0.8629 & 0.0393  & 0.0191 & 0.9820 & 0.0811 & 0.0413 & 0.9654  & 0.0682 & 0.0475 & 0.9512  \\
\textbf{CMCL} & 0.1628  & 0.1343  & 0.8594  & 0.0394   & 0.0203  & 0.9786  & 0.0815  & 0.0420  & 0.9598   & 0.1219  & 0.1116  & 0.9074 \\
\textbf{MMSAE} &0.0821  & 0.0657  & 0.9211  & 0.0347   & 0.0191  & 0.9846  & 0.0891  & 0.0413  & 0.9496   & 0.0460  & 0.0278  & 0.9699 \\
\textbf{PROSER} &0.0708  & 0.0555  & 0.9579  & 0.0387   & 0.0182  & 0.9834  & 0.0885  & 0.0406  & 0.9492   & 0.0693  & 0.0336  & 0.9653 \\
\textbf{HGM$^2$R} &0.1553  & 0.1613  & 0.8593  & 0.1452   & 0.0742  & 0.8996  & 0.2260  & 0.1265  & 0.8430   & 0.1006  & 0.0747  & 0.9305 \\
\midrule
\textbf{Ours} & \textbf{0.3013 } & \textbf{0.3202 } & \textbf{0.7048  } & \textbf{0.2811 } & \textbf{0.1377 } & \textbf{0.7316  } & \textbf{0.4471 } & \textbf{0.1914 } & \textbf{0.5936  } & \textbf{0.1277 } & \textbf{0.1117 } & \textbf{0.8994  } \\ 
\bottomrule
\end{tabular}
}
\label{table_4_2_2}
\end{center}
\end{table*}

\section{Experiments}
\label{section_4}
\subsection{Experimental Settings}
\label{section_4_1}
\textbf{OCMR Datasets.} We generate four open-set 3D cross-modal retrieval (OCMR) datasets, including OCAB, OCNT, OCES, OCMN, based on the public datasets ABO~\cite{collins2022abo}, NTU~\cite{chen2003visual}, ESB~\cite{jayanti2006developing}, and ModelNet40~\cite{wu20153d}, respectively. These datasets are split into seen and unseen categories, each object has three modalities including multi-view, voxel, and point cloud.

\noindent \textbf{Implemental Details.} In our experiment, we choose all three modalities of 3D objects. We set $\alpha=0.5$ for the hyper-parameters in Eq.~\ref{equation_3_3_5}, and $\tau=0.75$ in Eq.~\ref{equation_3_4_1}. The two modules are trained separately with 40 epochs on learning rate $lr=0.1$ and 120 epochs on $lr=0.001$, the random seed is fixed as 2022 for all experiments.

\subsection{Retrieval Performance}
\label{section_4_2}
\textbf{Compared Methods.} As no methods are specifically designed for the open-set 3D cross-modal retrieval to date, we refine the current state-of-the-art methods from two tasks for comparison: close-set 3D cross-modal retrieval (SDML~\cite{hu2019scalable}, CMCL~\cite{jing2021cross}, MMSAE~\cite{wu2019multi}), and open-set multi-modal recognition or retrieval (PROSER~\cite{zhou2021learning}, HGM$^2$R~\cite{feng2023hypergraph}).

\noindent \textbf{Evaluation Metrics.} For a fair comparison, we employ the commonly used retrieval metric, including Mean Average Precision (mAP), Normalized Discounted Cumulative Gain (NDCG), Average Normalized Modified Retrieval Rank (ANMRR), and the Precision-Recall Curve (PR-Curve). For the mAP and NDCG metric, higher scores are better. For the ANMRR metric, the lower score is better. We construct $6$ query-target types for cross-modal retrieval according to these three modalities, including Image2Point (I2P), Image2Voxel (I2V), Point2Image (P2I), Point2Voxel (P2V), Voxel2Image (V2I), and Voxel2Point (V2P). 

\begin{figure*}[!htbp]
    \subfigure[PR Curve on OCAB.]{
    	\begin{minipage}[t]{0.23\textwidth}
    		\centering
    		\includegraphics[bb=0 0 2159.28 1655.28, width=\textwidth]{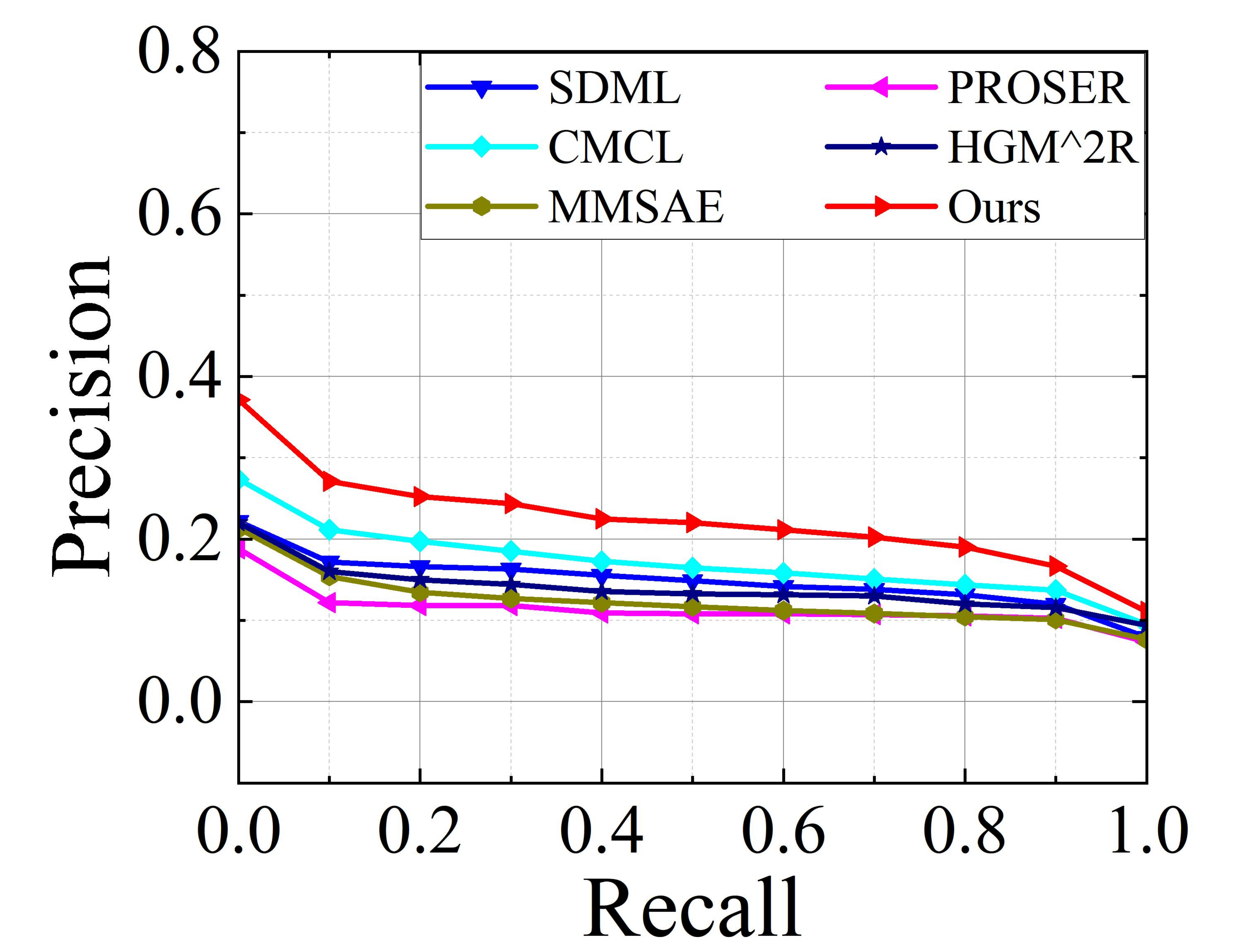}
    		\label{figure_4_2_1}
    	\end{minipage}
	}
    \subfigure[PR Curve on OCNT.]{
    	\begin{minipage}[t]{0.23\textwidth}
    		\centering
    		\includegraphics[bb=0 0 2159.28 1655.28, width=\textwidth]{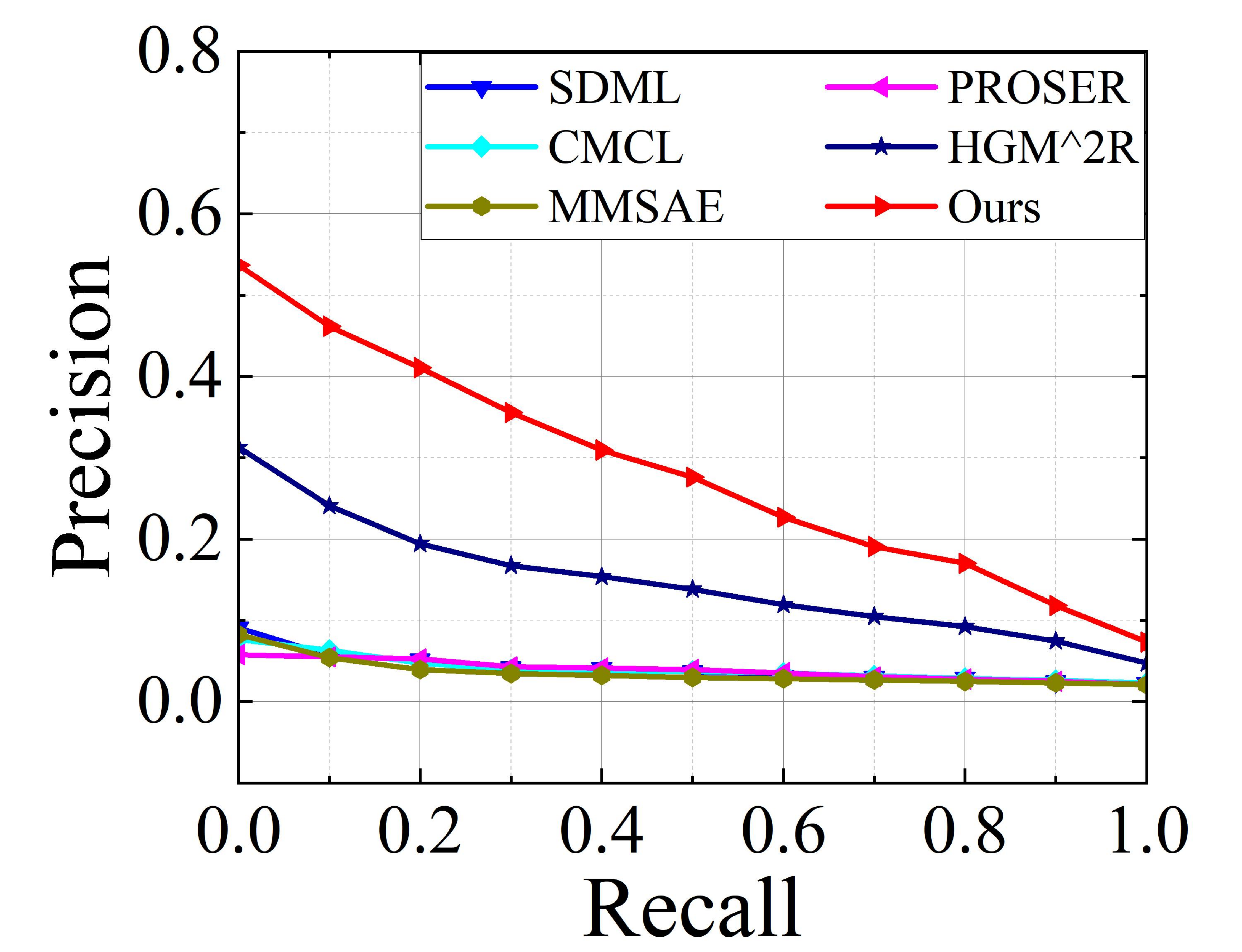}
    		\label{figure_4_2_2}
    	\end{minipage}
	}
    \subfigure[PR Curve on OCES.]{
    	\begin{minipage}[t]{0.23\textwidth}
    		\centering
    		\includegraphics[bb=0 0 2159.28 1655.28, width=\textwidth]{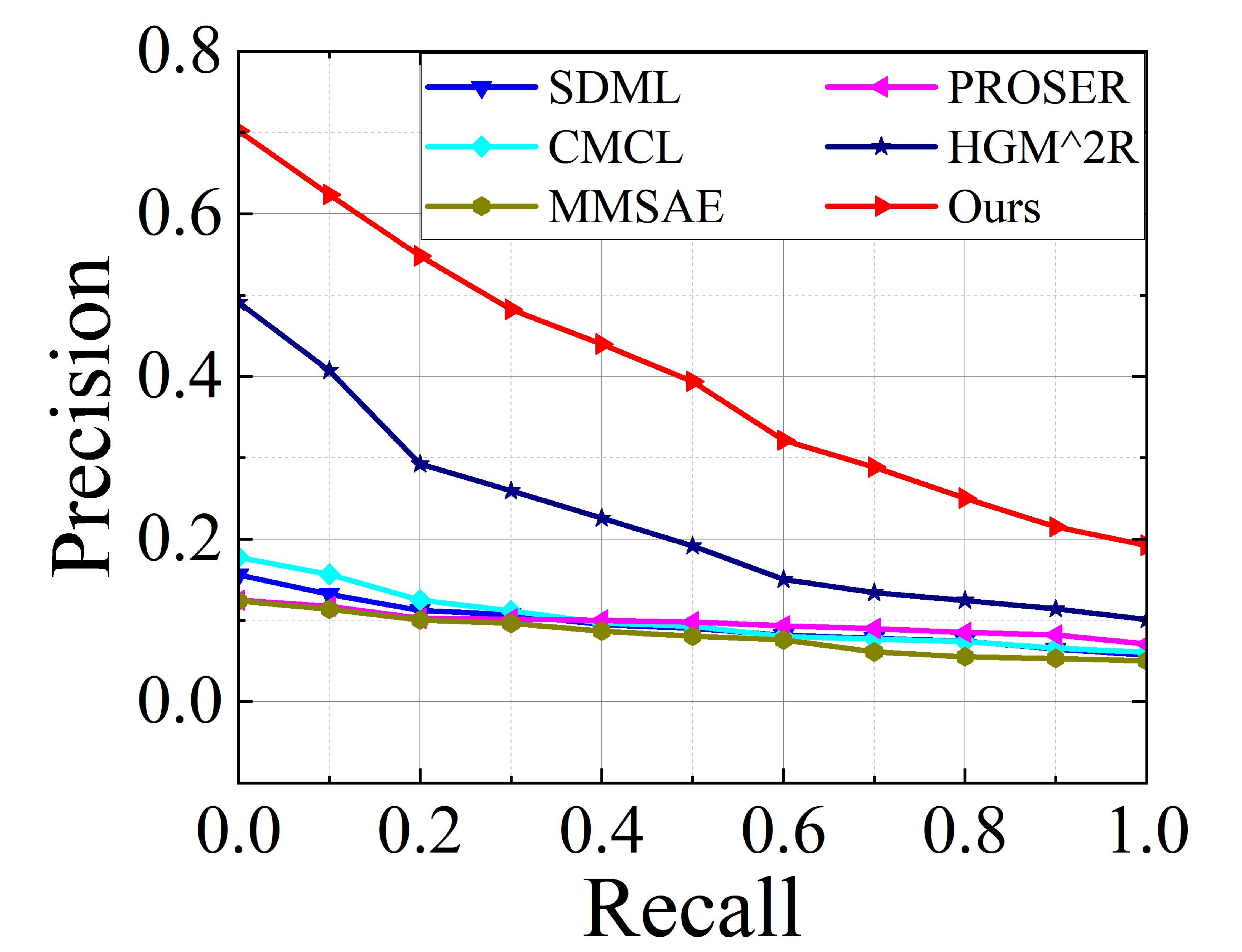}
    		\label{figure_4_2_3}
    	\end{minipage}
	}
    \subfigure[PR Curve on OCMN.]{
    	\begin{minipage}[t]{0.23\textwidth}
    		\centering
    		\includegraphics[bb=0 0 2159.28 1655.28, width=\textwidth]{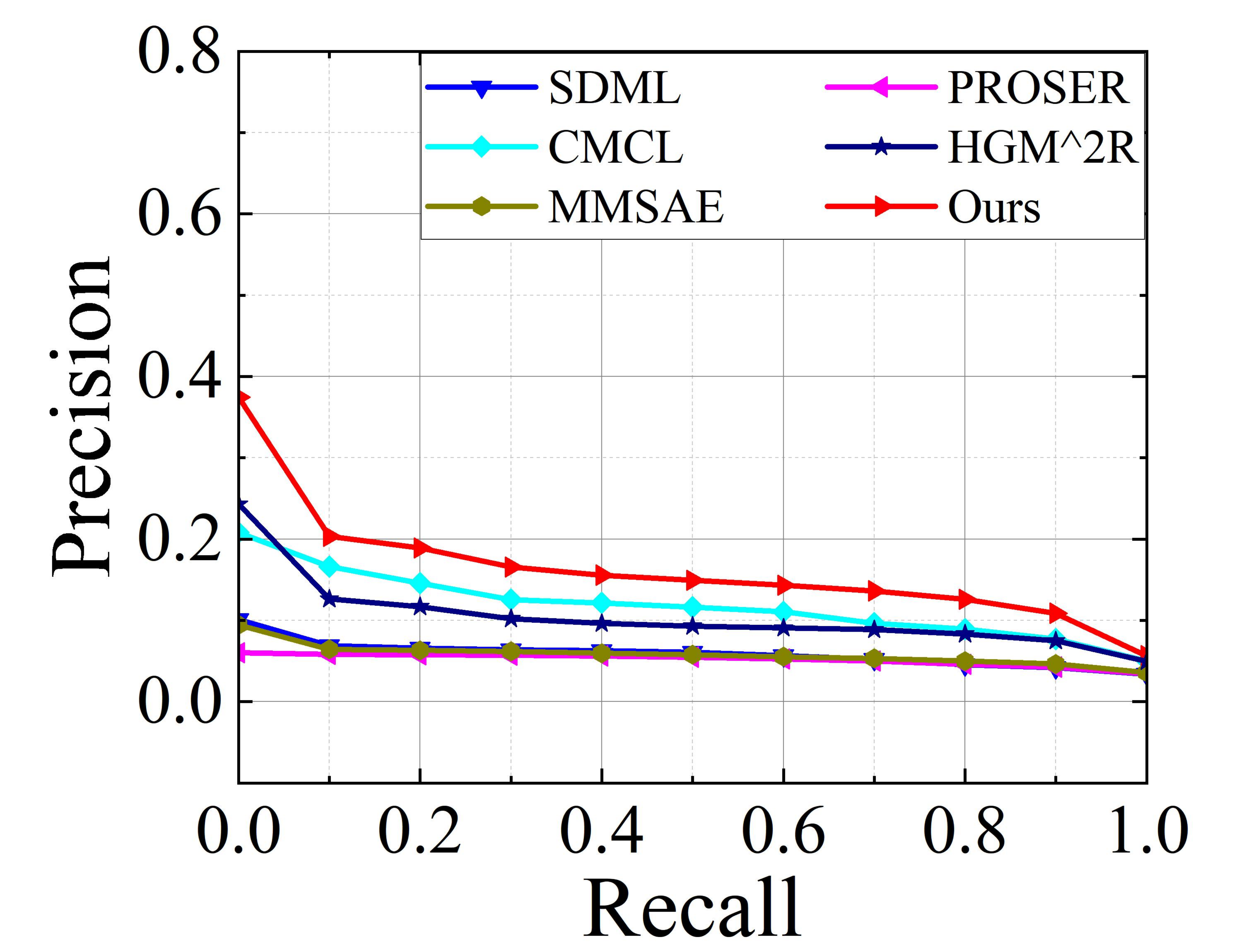}
    		\label{figure_4_2_4}
    	\end{minipage}
	}
	\caption{
	\label{figure_4_2}
	The precision-recall curves comparison of Image2Point retrieval on four datasets, respectively.}
\end{figure*}

\noindent \textbf{Comparison Analysis.} We evaluate open-set 3D cross-modal retrieval results on four datasets, quantitative results of {\ours} framework and other state-of-the-art methods are provided in Tab.~\ref{table_4_2_1} and Tab.~\ref{table_4_2_2}. Results show that the proposed method outperforms the other methods on all four datasets. We also provide the Precision-Recall (PR) Curve to evaluate the performance of the proposed {\ours} framework and other compared methods, as illustrated in Fig.~\ref{figure_4_2}. The larger area below the curve indicates better performance. From the results, we can observe that our method outperforms all other compared methods. The better performance indicates that by the residual-center embedding and hierarchical structure learning, the proposed method has the capability to overcome modality gaps while understanding the open-set categories.

\begin{table}[t]
\begin{center}
\caption{Ablation studies on OCNT dataset.}
\small
\scalebox{0.95}{
\begin{tabular}{ccccc}
\toprule
\multicolumn{1}{l}{} & \textbf{} & \textbf{mAP}$\uparrow$ & \textbf{NDCG}$\uparrow$ & \textbf{ANMRR}$\downarrow$ \\
\midrule
\multirow{2}{*}{\textbf{On RCE}} & \textbf{Direct Center} & 0.0362 & 0.0194 & 0.9801 \\
 & \textbf{Category Center} & 0.0433 & 0.0230 & 0.9772 \\
 \midrule
\multirow{4}{*}{\textbf{On HSL}} & \textbf{HSL w/o $\mathcal{E}_m$}  & 0.1511 & 0.1081 & 0.8626 \\
& \textbf{HSL w/o $\mathcal{E}_m$\&$\mathcal{E}_o$}  &0.1474 & 0.1084 & 0.8636 \\
 & \textbf{GCN-based HSL} &0.2575 & 0.1640 & 0.7374 \\
  & \textbf{MLP-based HSL} &0.1553 & 0.1101 & 0.8380 \\
 \midrule
\multicolumn{1}{l}{} & \textbf{RCE+HSL} & \textbf{0.2861} & \textbf{0.1585} & \textbf{0.7292}\\
\bottomrule 
\end{tabular}
}
\label{table_4_3}
\end{center}
\end{table}

\subsection{Ablation Study}
We conduct ablation studies to verify the effectiveness of the proposed modules. For the residual-center embedding module, we compare the proposed RCE with the \textit{Direct Center} and \textit{Category Center}. The \textit{Direct Center} denotes the network that use auto-encoder to generate the center embedding directly instead of residually, and \textit{Category Center} denotes the network that generates the category center rather than semantic center of each object. During the ablation of the hierarchical structure learning module, we compared the proposed HSL with naive structures (HSL w/o $\mathcal{E}_m$ and HSL w/o 
$\mathcal{E}_m$\&$\mathcal{E}_o$), where ``w/o'' denotes ``without''. We also replace the hypergraph-based correlation learning with MLP and GCN. As shown in Tab.~\ref{table_4_3}, the combination of RCE and HSL yields the best performance, substituting either the embedding or the learning approach is observed to lead to a decline in performance. These results demonstrate the proposed modules can effectively obtain the center-embedding of objects and generalize it to unseen categories.

\section{Conclusion}
In this paper, we propose the Structure-Aware Residual-Center Representation (\ours) framework for self-supervised open-set 3D cross-modal retrieval. We utilize the Residual-Center Embedding (RCE) for each object by nested auto-encoders to address the center deviation due to category distribution differences, rather than directly mapping them to the modality or category centerS. Besides, we construct a heterogeneous hypergraph structure based on hierarchical inter-modality, intra-object, and implicit-category correlations, and perform the Hierarchical Structure Learning (HSL) approach to leverage the high-order correlations among objects for generalization. Extensive experiments and ablation studies on four benchmarks demonstrate the superiority of our proposed framework compared to state-of-the-art methods.
\bibliographystyle{IEEEbib}
\bibliography{main}

\begin{thebibliography}{10}

\bibitem{jing2021cross}
Longlong Jing, Elahe Vahdani, Jiaxing Tan, and Yingli Tian,
\newblock ``{C}ross-modal {C}enter {L}oss for {3D} {C}ross-{M}odal {R}etrieval,''
\newblock in {\em CVPR}, 2021, pp. 3142--3151.

\bibitem{wang2017adversarial}
Bokun Wang, Yang Yang, Xing Xu, Alan Hanjalic, and Heng~Tao Shen,
\newblock ``{A}dversarial {C}ross-modal {R}etrieval,''
\newblock in {\em ACMMM}, 2017, pp. 154--162.

\bibitem{feng2023rono}
Yanglin Feng, Hongyuan Zhu, Dezhong Peng, Xi~Peng, and Peng Hu,
\newblock ``{RONO}: {R}obust {D}iscriminative {L}earning {W}ith {N}oisy {L}abels for {2D-3D} {C}ross-{M}odal {R}etrieval,''
\newblock in {\em CVPR}, 2023, pp. 11610--11619.

\bibitem{andrew2013deep}
Galen Andrew, Raman Arora, Jeff Bilmes, and Karen Livescu,
\newblock ``{D}eep {C}anonical {C}orrelation {A}nalysis,''
\newblock in {\em ICML}. PMLR, 2013, pp. 1247--1255.

\bibitem{wang2015deep}
Weiran Wang, Raman Arora, Karen Livescu, and Jeff Bilmes,
\newblock ``{O}n {D}eep {M}ulti-{V}iew {R}epresentation {L}earning,''
\newblock in {\em ICML}. PMLR, 2015, pp. 1083--1092.

\bibitem{vaze2021open}
Sagar Vaze, Kai Han, Andrea Vedaldi, and Andrew Zisserman,
\newblock ``{O}pen-{S}et {R}ecognition: {A} {G}ood {C}losed-{S}et {C}lassifier is {A}ll {Y}ou {N}eed?,''
\newblock {\em arXiv preprint arXiv:2110.06207}, 2021.

\bibitem{chen2021adversarial}
Guangyao Chen, Peixi Peng, Xiangqian Wang, and Yonghong Tian,
\newblock ``{A}dversarial {R}eciprocal {P}oints {L}earning for {O}pen {S}et {R}ecognition,''
\newblock {\em TPAMI}, vol. 44, no. 11, pp. 8065--8081, 2021.

\bibitem{feng2023hypergraph}
Yifan Feng, Shuyi Ji, Yu-Shen Liu, Shaoyi Du, Qionghai Dai, and Yue Gao,
\newblock ``{H}ypergraph-based {M}ulti-{M}odal {R}epresentation for {O}pen-{S}et {3D} {O}bject {R}etrieval,''
\newblock {\em TPAMI}, , no. 01, pp. 1--18, 2023.

\bibitem{feng2014cross}
Fangxiang Feng, Xiaojie Wang, and Ruifan Li,
\newblock ``{C}ross-{M}odal {R}etrieval with {C}orrespondence {A}uto{E}ncoder,''
\newblock in {\em ACMMM}, 2014, pp. 7--16.

\bibitem{gao2022hgnn+}
Yue Gao, Yifan Feng, Shuyi Ji, and Rongrong Ji,
\newblock ``{HGNN+}: {G}eneral {H}ypergraph {N}eural {N}etworks,''
\newblock {\em TPAMI}, vol. 45, no. 3, pp. 3181--3199, 2022.

\bibitem{collins2022abo}
Jasmine Collins, Shubham Goel, Kenan Deng, Achleshwar Luthra, Leon Xu, Erhan Gundogdu, Xi~Zhang, Tomas F~Yago Vicente, Thomas Dideriksen, Himanshu Arora, et~al.,
\newblock ``{ABO}: {D}ataset and {B}enchmarks for {R}eal-{W}orld {3D} {O}bject {U}nderstanding,''
\newblock in {\em CVPR}, 2022, pp. 21126--21136.

\bibitem{chen2003visual}
Ding-Yun Chen, Xiao-Pei Tian, Yu-Te Shen, and Ming Ouhyoung,
\newblock ``{O}n {V}isual {S}imilarity based {3D} {M}odel {R}etrieval,''
\newblock in {\em Computer graphics forum}. Wiley Online Library, 2003, pp. 223--232.

\bibitem{jayanti2006developing}
Subramaniam Jayanti, Yagnanarayanan Kalyanaraman, Natraj Iyer, and Karthik Ramani,
\newblock ``{D}eveloping an {E}ngineering {S}hape {B}enchmark for {CAD} {M}odels,''
\newblock {\em Computer-Aided Design}, vol. 38, no. 9, pp. 939--953, 2006.

\bibitem{wu20153d}
Zhirong Wu, Shuran Song, Aditya Khosla, Fisher Yu, Linguang Zhang, Xiaoou Tang, and Jianxiong Xiao,
\newblock ``{3D} {S}hapenets: {A} {D}eep {R}epresentation for {V}olumetric {S}hapes,''
\newblock in {\em CVPR}, 2015, pp. 1912--1920.

\bibitem{hu2019scalable}
Peng Hu, Liangli Zhen, Dezhong Peng, and Pei Liu,
\newblock ``{S}calable {D}eep {M}ultimodal {L}earning for {C}ross-{M}odal {R}etrieval,''
\newblock in {\em SIGIR}, 2019, pp. 635--644.

\bibitem{wu2019multi}
Yiling Wu, Shuhui Wang, and Qingming Huang,
\newblock ``{M}ulti-{M}odal {S}emantic {A}uto{E}ncoder for {C}ross-{M}odal {R}etrieval,''
\newblock {\em Neurocomputing}, vol. 331, pp. 165--175, 2019.

\bibitem{zhou2021learning}
Da-Wei Zhou, Han-Jia Ye, and De-Chuan Zhan,
\newblock ``{L}earning {P}laceholders for {O}pen-{S}et {R}ecognition,''
\newblock in {\em CVPR}, 2021, pp. 4401--4410.

\end{thebibliography}

\end{document}